\documentclass[aps,prl,superscriptaddress,reprint,aps,showpacs]{revtex4-1}

\usepackage{graphicx}
\usepackage{dcolumn}
\usepackage{bm}
\usepackage{color}
\usepackage{epstopdf}

\usepackage{ulem}
\usepackage{hyperref}
\usepackage{cleveref}

\newcommand{\figref}[1]{Fig.~\ref{#1}}
\newcommand{\Figref}[1]{Fig.~\ref{#1}}

\usepackage{tikz}

\begin{document}
\onecolumngrid
with kind permission from the American Physical Society, published as \href{https://link.aps.org/doi/10.1103/PhysRevB.97.115154}{Phys. Rev. B 97, 115154 (2018)}
\\\\
\twocolumngrid

\title{Inhibition of the photoinduced structural phase transition\\in the excitonic insulator Ta$_2$NiSe$_5$}

\author{Selene~Mor}
\email{mor@fhi-berlin.mpg.de}
\affiliation{Department of Physical Chemistry, Fritz-Haber-Institut der MPG, Faradayweg 4-6, 14195 Berlin, Germany}

\author{Marc~Herzog}
\affiliation{Department of Physical Chemistry, Fritz-Haber-Institut der MPG, Faradayweg 4-6, 14195 Berlin, Germany}
\affiliation{Institute for Physics and Astronomy, University of Potsdam, Karl-Liebknecht-Str. 24-25, 14476 Potsdam, Germany}

\author{Johannes~Noack}
\affiliation{Department of Inorganic Chemistry, Fritz-Haber-Institut der MPG, Faradayweg 4-6, 14195 Berlin, Germany}

\author{Naoyuki~Katayama}
\affiliation{Department of Physical Science and Engineering, Nagoya University, 464-8603 Nagoya, Japan}

\author{Minoru~Nohara}
\affiliation{Research Institute for Interdisciplinary Science, Okayama University, Okayama 700-8530, Japan}

\author{Hide~Takagi}
\affiliation{Max Planck Institute for Solid State Research, 70569 Stuttgart, Germany}
\affiliation{Department of Physics, University of Tokyo, 113-8654 Tokyo, Japan}

\author{Annette~Trunschke}
\affiliation{Department of Inorganic Chemistry, Fritz-Haber-Institut der MPG, Faradayweg 4-6, 14195 Berlin, Germany}

\author{Takashi~Mizokawa}
\affiliation{Department of Applied Physics, Waseda University, Tokyo 169-8555, Japan}

\author{Claude~Monney}
\affiliation{Department of Physics, University of Fribourg, 1700 Fribourg, Switzerland}

\author{Julia~St\"{a}hler}
\affiliation{Department of Physical Chemistry, Fritz-Haber-Institut der MPG, Faradayweg 4-6, 14195 Berlin, Germany}

\date{\today}

\begin{abstract}

Femtosecond time-resolved mid-infrared reflectivity is used to investigate the electron and phonon dynamics occurring at the direct band gap of the excitonic insulator Ta$_2$NiSe$_5$ below the critical temperature of its structural phase transition. We find that the phonon dynamics show a strong coupling to the excitation of free carriers at the $\Gamma$ point of the Brillouin zone. The optical response saturates at a critical excitation fluence $F_C = 0.30~\pm~0.08$~mJ/cm$^2$ due to optical absorption saturation. This limits the optical excitation density in Ta$_2$NiSe$_5$ so that the system cannot be pumped sufficiently strongly to undergo the structural change to the high-temperature phase. We thereby demonstrate that Ta$_2$NiSe$_5$ exhibits a blocking mechanism when pumped in the near-infrared regime, preventing a nonthermal structural phase transition.

\end{abstract}


\maketitle
\begin{center}
\textbf{I. INTRODUCTION}
\end{center}
The interplay of electronic and structural degrees of freedom crucially determines both the ground- and excited-state properties of solids.
Understanding how the electron and lattice subsystems couple and if it is possible to control them independently is of great interest both fundamentally and for applications. In equilibrium, this can be difficult as all the relevant properties manifest concurrently. Optical perturbation of the system allows the disentanglement and, potentially, the selective control of the subsystems. However, the relevant processes occur on ultrafast time scales, which makes the detection challenging. Examples are the generation of spin-density-wave order in the pnictide compound BaFe$_2$As$_2$ by coherent lattice vibrations \cite{Kim12} or the attempts to induce superconductivity via selective phonon pumping in high-$T_C$ superconductors like La$_{1.675}$Eu$_{0.2}$Sr$_{0.125}$CuO$_{4}$ \cite{Fausti11}. By driving the system out of equilibrium new transient states can be accessed which differ from those in the equilibrium phase diagram. In this case, the electronic and structural subsystems may exhibit transiently decoupled evolutions, as in the case of the instantaneous metalization of the strongly correlated electron system VO$_2$ \cite{Morrison14, Wall2012, Wegk2014,Wegk15} or a system can switch to a metastable hidden state, as observed in the layered chalcogenide 1$T$-TaS$_{2}$ \cite{Stoj2014}. Often it would be helpful if one of the subsystems remained unaffected by the photoexcitation, allowing the unperturbed investigation of the other one and, specifically, of the leading mechanisms of the dynamics and the equilibrium properties of a specific phase.

The small-band-gap semiconductor Ta$_2$NiSe$_5$ (TNS) belongs to the family of two-dimensional chalcogenides. It shows an intralayer quasi-one-dimensional crystal structure with parallel chains of Ta and Ni atoms \cite{Suns1985}. A gap in the band structure occurs at the $\Gamma$ point of the Brillouin zone, such that an exciton can form as a result of a low-energy excitation across the atomic chains  \cite{Kane2013}. Thus, excitonic properties are likely to be connected to any structural change involving the interchain distance. Below the critical temperature $T_C=328$~K, the Ta-Ni separation shortens due to a second-order crystallographic phase transition without any signature of a charge-density wave, consistent with its direct band gap \cite{DiSa1986,Cana1987}. At the same $T_C$, the band gap widens \cite{Waki2009}, and it has been shown that the combined structural and electronic phase transition is consistent with exciton condensation \cite{Kane2013,Zenk2014,Seki2014,Ejim2014,Lu17}. Such an excitonic insulator (EI) phase has been predicted theoretically for small-band-gap semiconductors or semimetals in which the exciton binding energy is larger than the electronic band gap \cite{Mott1961,Kohn1967,Jero1967}. In this case, excitons form spontaneously and can condense into a new insulating ground state. Currently, the ultrafast dynamics of the EI phase are attracting significant interest, particularly with regard to their relevance to the photoinduced EI gap dynamics and their interplay with different degrees of freedom \cite{Gol16,Mura17}. In our previous work, we performed time- and angle-resolved photoelectron spectroscopy (trARPES) of the low temperature (LT) phase of TNS and demonstrated a massive depopulation of the upper valence band (VB) around $\Gamma$ by near-infrared (1.55~eV) pump pulses which causes photon absorption saturation at a critical fluence of $0.2$~mJ/cm$^2$ \cite{Mor2017}. Further, the electronic band gap transiently widens above the saturation fluence due to an increase of the exciton condensate density. We demonstrated that photoexcitation does not induce the electronic phase transition to the high-temperature (HT) semiconducting phase and that the system rather evolves into an excited EI state. Despite all the evidence that the electronic and the lattice subsystems are entwined, it is still unclear how the relevant transitions are connected. Moreover, it is an open question whether the structural phase transition can be photoinduced in the absence of the EI-semiconductor phase transition. 

In the present study, we capture the nonequilibrium lattice and electron dynamics in the LT phase of TNS by means of time-resolved midinfrared (MIR) reflectivity measurements. The dynamics of the excited coherent optical phonons, whose frequencies are compared with Raman spectra at the respective temperatures, reveal that the structural phase transition cannot be driven nonthermally by near-infrared  (1.55 eV) photoexcitation. The carrier dynamics are directly monitored by the incoherent optical response. Both the coherent and incoherent reflectivity signals follow the fluence dependence of the photoinduced hole population of the VB at $\Gamma$ \cite{Mor2017}, including the saturation behavior at a comparable excitation density $F_C = 0.30~\pm~0.08$~mJ/cm$^2$. We thus reveal that the electron and phonon dynamics are strongly coupled and we argue that a nonthermally photoinduced structural phase transition in TNS is inhibited by complete absorption saturation of 1.55 eV pump pulses below the energy threshold for the structural change.

\begin{center}
\textbf{II. EXPERIMENTAL DETAILS}
\end{center}
Single crystals of TNS were grown by chemical vapor transport. The Raman spectra are recorded using a 532-nm laser source and a TriVista Raman Microscope System TR557 (S\&I GmbH) equipped with a 750-mm monochromator (Princeton Instruments). Because of the experimental geometry and the reduced sample dimensions, direct temperature measurements on the sample are difficult. For this reason, we measure the temperature of the sample holder, resulting in an independently evaluated discrepancy of approximately 15 K.  Time-resolved experiments are performed using a Coherent RegA 9040 laser system which generates 40-fs laser pulses at 1.55~eV at a repetition rate of 40~kHz. Half of the laser output drives a two-stage noncolinear optical parametric amplifier generating MIR pulses at 300~meV \cite{Brad2011}. The reflected MIR light is detected by a liquid-nitrogen-cooled HgCdTe detector using a lock-in amplifier. The transient MIR reflectivity change induced by the photoexcitation with 1.55~eV pump pulses is measured after variable time delays.
 
\begin{center}
\textbf{III. RESULTS AND DISCUSSION}
\end{center}
\begin{figure}
\includegraphics[width=0.9\columnwidth]{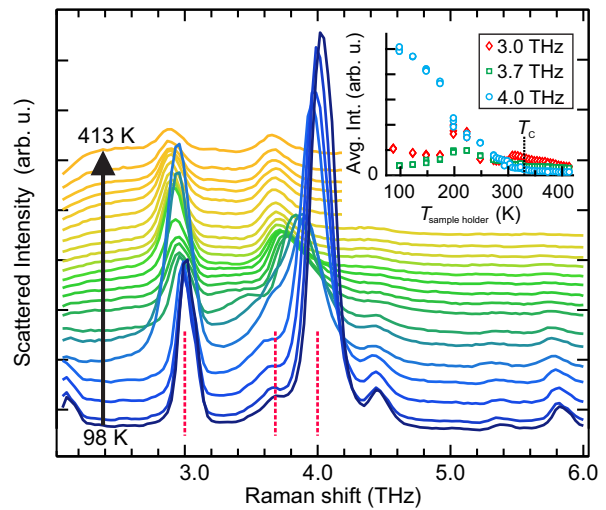}
\caption{(a) Temperature-dependent Raman spectra of TNS. Raman modes at 3.0, 3.7, and 4.0~THz are indicated by the dashed lines. Inset: Averaged intensity of the Raman peaks as a function of the temperature.
}
\label{fig1}
\end{figure}

The Raman spectra at different temperatures are shown in \figref{fig1}. At low temperatures (blue curves) we observe three phonon modes at 3.0, 3.7 and 4.0~THz while only the two lower-frequency modes are present above $T_C$ (yellow curves). The inset shows the averaged intensity of each peak as a function of temperature: the intensity of the 3.0- and 3.7-THz modes remains rather constant, while that of the 4.0-THz mode drops continuously as the temperature increases up to $T_C$. The decrease in the number of phonon modes from three to two at higher temperatures is consistent with an increase in the lattice symmetry during the structural phase transition \cite{Wall2012}. While the 3.0- and 3.7-THz modes are insensitive to the crystal symmetry change, the temperature evolution of the 4.0-THz mode intensity reflects the second-order character of the transition \cite{DiSa1986,Cana1987}. We prove that the structural transition to the HT phase is successfully driven upon heating and that it can be tracked by verifying the disappearance of the 4-THz mode, a fingerprint of the LT phase.

\begin{figure}
\includegraphics[width=0.9\columnwidth]{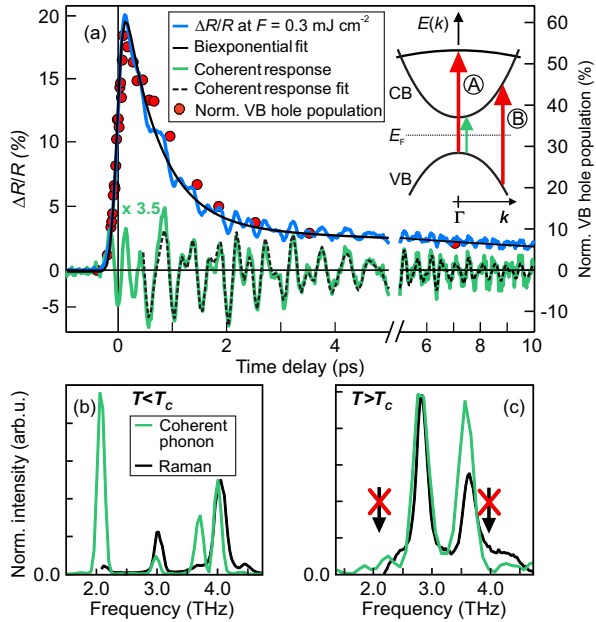}
\caption{(a) MIR reflectivity change in the LT phase as a function of time delay (blue and green curves) and relative fit curves (solid and dashed black).  The trARPES intensity (red circles) of the upper VB at $\Gamma$ is shown for comparison, modified from \cite{Mor2017}. Inset: Scheme of the electronic band structure of TNS. (b) and (c) Fourier transform of the coherent response (green curve) and Raman spectra (black curve) of LT and HT phases.}
\label{fig2}
\end{figure}
We turn to the time-domain experiments to gain insight into the non-equilibrium dynamics. As sketched in the inset of \Figref{fig2}(a), 1.55~eV pump photons excite two vertical electronic transitions from the upper VB: (A) into flat unoccupied \textit{d}~bands at $\Gamma$ and (B) into the lowest conduction band (CB) at larger \textit{k} vectors. Both excitation channels mostly involve electron transfer from Ni 3$d$ to Ta 5$d$ orbitals \cite{Kane2013}, with likely excitation of phonons. With 300 meV energy (green arrow), the probe photons are predominantly sensitive to carrier dynamics around the VB and the CB extrema at $\Gamma$ and as such, to the low-energy physics of the phase transition in TNS \cite{Seki2014,Waki2012}. The blue trace in \figref{fig2}(a) shows the transient relative change of the MIR reflectivity in the LT phase at 120~K for an excitation density of 0.30~mJ/cm$^2$. The signal consists of an incoherent optical response superimposed by an oscillatory part. Upon photoexcitation, the incoherent response increases abruptly by several percent and exhibits a biexponential decay (solid black curve) with time constants on the order of 1 and 15~ps, respectively \footnote{The observed time constants increase linearly with increasing pump fluence (not shown).}. We note that this signal coincides with the dynamics of the hole population in the photoexcited upper VB at $\Gamma$ (red circles) probed by trARPES (data taken from \cite{Mor2017}), thus showing that the incoherent optical response directly monitors the carrier dynamics in the vicinity of $E_F$ launched by the near-infrared photoexcitation.

The pure oscillatory response (green cure) is due to the excitation of coherent optical phonons and is obtained by subtracting the biexponential fit of the incoherent response from the data. The corresponding phonon spectra of the LT phase at 120~K [\figref{fig2}(b)] and of the HT phase at 330~K [\figref{fig2}(c)] are extracted by fast Fourier transformation. They agree well with the Raman spectra (black curve) at comparable temperatures. An additional phonon mode at 2.1 THz is present solely in the LT phase, whose frequency was not accessible in the Raman experiments. The coherent optical signal in the LT phase can then be accurately modeled by a sum of four damped oscillations as shown by the dashed black trace in Fig. 2(a). We note that the disappearance of the 2.1- and 4.0-THz phonon modes above $T_C$ in the time-domain data confirms that transient MIR reflectivity experiments are sensitive to the transition to the HT structural phase if it occurs. 

\begin{figure}
\includegraphics[width=\columnwidth]{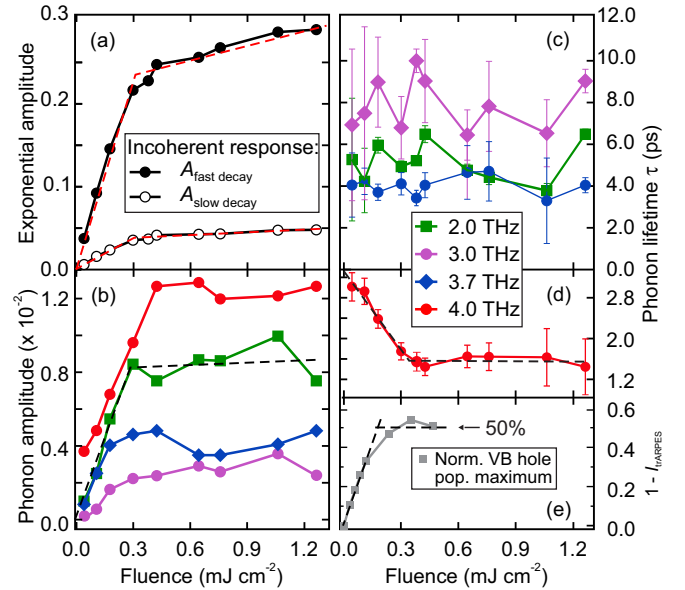}
\caption{(a)-(d) Fit results of the incoherent and coherent responses in the LT phase at $T=120$~K for various excitation fluences and relative broken line fit curves (dashed red and black): (a) Amplitudes of fast and slow exponential decay of the incoherent response, (b) amplitudes of the phonon modes, (c) fluence-independent lifetimes of the 2.1-, 3.0- and 3.7-THz phonon modes, and (d) fluence-dependent lifetime of the LT phonon mode at 4.0~THz. (e) Transient hole population maxima in the upper VB at $\Gamma$. Adapted from \cite{Mor2017}.
}
\label{fig3}
\end{figure}
\figref{fig3}(a) summarizes the fluence dependence of the incoherent MIR optical response below $T_C$: At low fluences, the amplitudes of the exponential functions fitted to the incoherent responses increase linearly with increasing pump fluence. However, the dependence is strongly reduced at $F_C = 0.30~\pm~0.08$~mJ/cm$^2$, as extracted by the broken line fit (dashed red line). We compare this fluence dependence to that of the hole population of the upper VB at $\Gamma$ shown in \figref{fig3}(e) (data taken from \cite{Mor2017}), as this provides the main initial state for the MIR resonances in the reflectivity measurements. The VB hole population forms via the excitation mechanism (A) [see inset in \figref{fig2}(a)]. The effect scales linearly with increasing pump fluence until half occupancy occurs at approximately $0.2$~mJ/cm$^2$. At higher fluences, the hole population remains constant due to absorption saturation of pump photons in this two-level system, as discussed in detail in \cite{Mor2017} and corroborated by a fluence-dependent change in reflectivity at 1.55 eV photon energy (not shown). The saturation fluence is comparable to the $F_C$ value at which the MIR incoherent optical response undergoes an abrupt change. Both critical values refer to the incident, not the absorbed fluence \footnote{The absorbed energy at the critical fluence can be estimated using optical constants and specific heat from Refs. \cite{OkamuraPrivate,Lar17}, respectively and amounts to a maximal temperature increase of approximately 50 K, well below $T_C$.}. The discrepancy between them is then consistent with the different sample volumes probed by optical and photoemission experiments, being bulk and surface sensitive, respectively. We conclude that the fluence dependence of the MIR incoherent optical response changes above $F_C$ because of optical absorption saturation at $\Gamma$. However, the MIR optical response is also sensitive to charge-carrier dynamics at $\Gamma$ that are caused by excitation mechanism (B). As the incoherent response amplitude does not completely saturate above $F_C$ but continues to increase with a smaller slope, we infer that this excitation channel persists at high fluences. 

The fluence dependence of the coherent optical response allows us to investigate the structural dynamics photoinduced in the LT phase of TNS. If the photoinduced structural phase transition occurs abruptly, the amplitudes of the 2.1- and 4.0-THz phonons are expected to decrease at higher fluences and eventually vanish once the energy threshold for the lattice change is reached. In the case of finite phonon amplitudes, the phase transition may still occur nonthermally: the lifetime of the 2.1- and 4.0-THz LT phonon modes is then a lower boundary of the persistence of the LT crystal phase following the photoexcitation \cite{Morrison14}. Finally, we point out that several systems undergoing structural changes \cite{Yusu08, Schae14,Mohr2011} exhibit pronounced phonon frequency softening across the photoinduced structural phase transition. However, TNS should differ from these cases as no significant frequency shift is observed between the coherent phonon spectra above and below $T_C$ (see \figref{fig1}). 

The best-fit parameters of the coherent optical response are reported in Figs. 3(b)-3(d) as a function of pump fluence. We observe that the coherent phonon amplitudes [\figref{fig3}(b)] \textit{increase} linearly at low fluences and reach full saturation at $F_C$ \footnote{The relative amplitudes of the coherent phonons depend partially on the time resolution (of 170~fs). The actual relative amplitude of the high-frequency phonons should be higher.}. We thus infer that the transition to the HT structural phase does not occur within the pulse duration. As these fluence dependencies resemble that of the VB hole buildup at $\Gamma$ [\figref{fig3}(e)], we deduce that the phonon excitation occurs via excitation mechanism (A). 

In order to elucidate if the transition is driven nonthermally at all, we look at the lifetime of the excited phonon modes. The lifetimes of the 2.1-, 3.0-, and 3.7-THz modes [\figref{fig3}(c)] range from 4 to 8~ps and are independent of the pump fluence within experimental accuracy. Thus, electron-phonon scattering, which is strongest at earlier delay times \cite{Pet97}, cannot be the main decay mechanism for these lattice vibrations. On the contrary, the lifetime of the LT phonon mode at 4.0~THz [\figref{fig3}(d)] shortens linearly as the fluence is increased towards $F_C$, but for $F>F_C$, it remains constant at 1.57 $\pm$ 0.11~ps. Such a rather long lifetime of both the LT phonon modes at 2.1 and 4.0~THz even for high-excitation densities reveals a persistent LT crystal symmetry and the \textit{absence of a nonthermally induced transition} to the HT crystallographic phase in TNS. Both the linear dependence below $F_C$ and the subsequent saturation strongly suggest scattering with free carriers at $\Gamma$ to be the rate-determining factor of the 4.0-THz mode lifetime. 
Finally, we note that the two LT phonon modes at 2.1 and 4.0~THz never show a softening to zero energy (not shown). 

These results show that the structural dynamics of TNS in the LT phase are strongly connected to the free-carrier dynamics photoexcited at $\Gamma$ by near-infrared photons. This strong coupling of the two subsystems leads, in combination with the absorption saturation at $\Gamma$, to an inhibition of the photoinduced structural phase transition in TNS before the required energy threshold can be reached upon near-infrared photoexcitation. It should be noted that the energy absorbed at the saturation threshold exceeds the estimated exciton condensation energy of TNS of U$_{cond}$ $\approx$ 7 J/cm$^{3}$ \footnote{This is calculated as the energy difference in the occupied electronic band structure between the normal phase and the EI phase and has been estimated from the ARPES data in \cite{Mor2017} considering the topmost valence band.}. We conclude that some of this excess energy might be lost in scattering processes with phonons, while a large part contributes to exciting more electrons and holes which eventually enhance the exciton condensate density. This behavior is very different compared to that of other symmetry-broken-state materials in which a saturation of the optical response occurs at the condensate melting point \cite{Stoj11}.

\begin{figure}
\includegraphics[width=0.6\columnwidth]{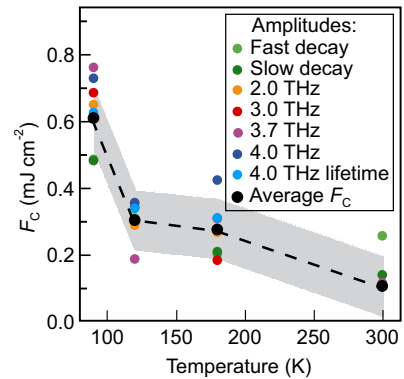}
\caption{Temperature dependence of $F_C$ for all observables.}
\label{fig4}
\end{figure}

In many materials \cite{Liu11,Por14,Lou16,Tao16}, the fluence threshold of a photoinduced phase transition depends strongly on temperature, as less energy is necessary to drive the~transition close to $T_C$. To test the validity range of the blocking mechanism in TNS, we performed temperature-dependent experiments. A saturation threshold persists in both the incoherent and coherent optical responses between 90 and 300~K and the relevant critical fluences are reported in \figref{fig4}. For every observable, the breaking point of the broken line fit is plotted as a function of temperature. All show the same trend and decrease by a factor of 3 when the sample is heated from 90 to 300~K. These findings corroborate once more that similar physics governs all saturation behaviors, which is summarized by the temperature dependence of the average $F_C$ (black circles; the shaded area is the average error). 

The decreasing $F_C$ as the system approaches $T_C$ is indicative of an effective lowering of the pump absorption saturation threshold, which could have different origins. The pump absorption coefficient may vary with temperature as the VB at $\Gamma$ shifts upon heating \cite{Waki2009,Waki2012,Seki2014,Mor2017}. However, this hypothesis is ruled out by the dielectric function obtained from reflectivity \cite{OkamuraPrivate} and ellipsometry \cite{Lar17} measurements, which shows the penetration depth for the near-infrared pump photons does not vary as a function of temperature. One plausible explanation would be the assignment of the change in $F_C$ to temperature-dependent variations of the free-carrier scattering rate. At high temperature, more scattering events with, e.g., thermally activated phonons and hot carriers can depopulate the excited states more rapidly and reenable the absorption of pump photons within the pulse duration, thus increasing $F_C$. However, in TNS the partial melting of the exciton condensate towards $T_C$ results in free carriers enhancing the Coulomb screening. As a consequence, the scattering probability from the excited states would be reduced, decreasing $F_C$. We observe that the threshold for the optical absorption saturation effectively decreases at higher temperatures. This finding suggests that enhanced screening of the Coulomb interaction dominates the early dynamics at elevated temperatures and ensures self-protection of the LT crystal phase of TNS against a photoinduced phase transition \textit{even close to $T_C$}.

\begin{center}
\textbf{IV. CONCLUSION}
\end{center}
In conclusion, by means of time-resolved MIR reflectivity we have successfully tracked the coupled electron and phonon dynamics photoinduced in the LT phase of the layered chalcogenide TNS. We have revealed that upon near-infrared photoexcitation the system does not switch to the HT structural phase nonthermally, but rather evolves to a thermally inaccessible transient excited state with unchanged crystal symmetry. As the structural transition does not take place upon photoexcitation of the LT phase, we can assign all ultrafast phenomena of the electronic subsystem to carrier and exciton dynamics. This result is complementary and in excellent agreement with the photoinduced EI band gap enhancement unveiled by previous trARPES experiments \cite{Mor2017}. We demonstrate that the blocking mechanism of the structural phase transition relies on the optical absorption saturation occurring at $\Gamma$ and occurs also for elevated temperatures close to $T_C$. This study not only unveils the interplay between the structural and the electronic subsystems in TNS but also represents an exemplary case of how the electronic subsystem can dynamically protect a structural phase against a photoinduced phase transition. 

\begin{center}
\textbf{ACKNOLEDGMENTS}
\end{center}
J.S. acknowledges the financial support from the Deutsche Forschungsgemeinschaft through Sfb951. C.M. acknowledges the financial support from the SNSF Grants No. PZ00P2\_154867 and No. PP00P2\_170597.

\bibliography{Publications-2016_TNS_phonons}

\end{document}